\documentclass[12pt]{article}
\usepackage[dvips]{color}
\usepackage{epsfig}
\usepackage{amsmath}
\usepackage{graphicx}

\begin{document}
\title{\bf Effect of Varying Bulk Viscosity on Generalized Chaplygin Gas}
\author{{H. Saadat \thanks{Email: hsaadat2002@yahoo.com} and B. Pourhassan \thanks{Email: b.pourhassan@umz.ac.ir}}\\
{\small {\em Department of  physics, Shiraz Branch, Islamic Azad
University, Shiraz, Iran}}} \maketitle
\begin{abstract}
\noindent In this paper, viscous generalized Chaplygin gas as a
model of dark energy considered. We assume non-constant bulk viscous
coefficient and study dark energy density. We consider several cases
of density-dependent viscosities.
We find that, in the special case, the viscous generalized Chaplygin gas is corresponding to modified Chaplygin gas.
\end{abstract}
\section{Introduction}
Dark energy and any studies about dark universe are important in
theoretical physics and cosmology. In order to understand nature of
dark universe one can investigate time-dependent density of dark
energy [1]. There are several models to describe dark energy [2-16].
We are interested to the cases of Chaplygin gas (CG) as a model for
dark energy [17-24], which is extended to generalized Chaplygin gas
(GCG) for observational agreements [25-29]. Also, it is possible to
extend GCG to the modified Chaplygin gas (MCG) [30], or modified
cosmic Chaplygin gas (MCCG) [31]. Moreover, we know that the bulk
viscosity plays an important role in cosmology [32, 33, 34]. The
idea that Chaplygin gas may has viscosity first proposed by the Ref.
[35] and then developed by the Refs. [36-41]. In the Refs. [31, 36]
we studied viscous modified cosmic Chaplygin gas and viscous
modified Chaplygin gas, and calculated time-dependent energy
density. In the Ref. [38] we studied viscous Chaplygin gas in
non-flat FRW universe. In these works we considered special case of
$\alpha=0.5$, so extension of viscous generalized Chaplygin gas for
arbitrary $\alpha$ performed in the Ref. [39]. In all cases the
viscous parameter considered as a constant. In the Ref. [42] it is
pointed that the viscous coefficient may be considered proportional
to powers of density ($\zeta\propto\rho^{n}$). So, in this work we
study effect of non-constant bulk viscosity on generalized Chaplygin
gas.
\section{Viscous generalized Chaplygin gas cosmology}
The generalized Chaplygin gas which unified dark energy and dark
matter described by the following equation of state,
\begin{equation}\label{s1}
p=-\frac{A}{\rho^{\alpha}},
\end{equation}
where $A$ is a positive constant and $0<\alpha\leq1$, so $\alpha=1$
gives Chaplygin gas equation of state. As we know the
Friedmann-Robertson-Walker (FRW) universe is described by the
following metric,
\begin{equation}\label{s2}
ds^2=-dt^2+a(t)^2(dr^2+r^{2}d\Omega^{2}),
\end{equation}
where $d\Omega^{2}=d\theta^{2}+\sin^{2}\theta d\phi^{2}$, and $a(t)$
represents the scale factor. We assume that the universe is filled
with the generalized Chaplygin gas with equation of state (1) and
neglect contribution of other components. Also,
\begin{equation}\label{s3}
\bar{p}=p-3\zeta(\rho) H,
\end{equation}
is the total pressure which involves the proper pressure $p$, given
by equation of state (1), bulk viscosity coefficient $\zeta(\rho)$
and Hubble expansion parameter $H=\dot{a}/a$. Also conservation
equation is given by,
\begin{equation}\label{s4}
\dot{\rho}+3H(\bar{p}+\rho)=0.
\end{equation}
Now, one can obtain the following field equations,
\begin{equation}\label{s5}
H^{2}=\frac{\rho}{3},
\end{equation}
and
\begin{equation}\label{s6}
\dot{H}+H^{2}=-\frac{\rho}{6}-\frac{\bar{p}}{2},
\end{equation}
where dot denotes derivative with respect to cosmic time $t$. By
using the equation of state (1) and total pressure in the
energy-momentum conservation formula (4) one can obtain the
following differential equation,
\begin{equation}\label{s7}
\dot{\rho}+3H \left(\rho-\frac{A}{\rho^{\alpha}}-3\zeta(\rho)
H\right)=0.
\end{equation}
Main goal of this paper is considering $\rho$-dependent viscous
coefficient instead of constant one. If we assume $\zeta(\rho)=0$,
then the energy density is obtained as the following,
\begin{equation}\label{s8}
\rho=\left[A+\frac{C}{a^{3(1+\alpha)}}\right]^{\frac{1}{1+\alpha}},
\end{equation}
where $C$ is an integration constant.
\section{$\rho$-dependent bulk viscosity}
It is interesting to choose the bulk viscous coefficient as the
following $\rho$-dependent expression,
\begin{equation}\label{s9}
\zeta(\rho)=\zeta_{0}\rho^{n},
\end{equation}
where $\zeta_{0}$ is a constant. In that case the equation (7) may
be written as the following,
\begin{equation}\label{s10}
\dot{\rho}+\sqrt{3}\sqrt{\rho}
\left(\rho-\frac{A}{\rho^{\alpha}}\right)-3\zeta_{0}\rho^{n+1}=0.
\end{equation}
Our main goal is solving the equation (10) to obtain time-dependent
density. Then, we are able to discuss evolution of scale factor and
Hubble expansion parameter.\\
Special case of $n=-\alpha-\frac{1}{2}$ yields to the following
expression in terms of the hypergeometric function,
\begin{equation}\label{s11}
t=\frac{2\sqrt{3}}{3}\rho^{-\frac{1}{2}} {_{2}F_{1}}\left(1,
\frac{1}{2(1+\alpha)}; 1+\frac{1}{2(1+\alpha)}; f(\rho; \alpha, A,
\zeta_{0})\right).
\end{equation}
In the Fig. 1 we show behavior of energy density in terms of time
for $n=-0.6$, $n=-1$ and $n=-1.4$. This case corresponds to negative
$n$ which may be unphysical, because we expect that increasing
density, increased viscosity. But for the negative $n$ we find that
increasing density decreased viscosity. Therefore, we will consider
another special cases with positive $n$.

\begin{figure}[th]
\begin{center}
\includegraphics[scale=.4]{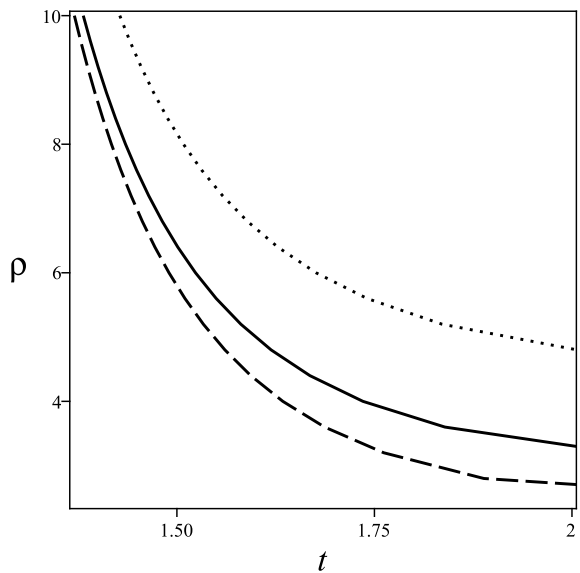}
\caption{Time-dependent dark energy density for
$n=-\alpha-\frac{1}{2}$ with $\alpha=0.1$ (dotted line),
$\alpha=0.5$ (solid line), $\alpha=0.9$ (dashed line).}
\end{center}
\end{figure}

In the Fig. 2 we draw behavior of energy density in terms of time
for $\alpha=0.5$ and some positive value of $n$. We find that
$n\leq0.2$ is essential condition to avoid singularity.

\begin{figure}[th]
\begin{center}
\includegraphics[scale=.4]{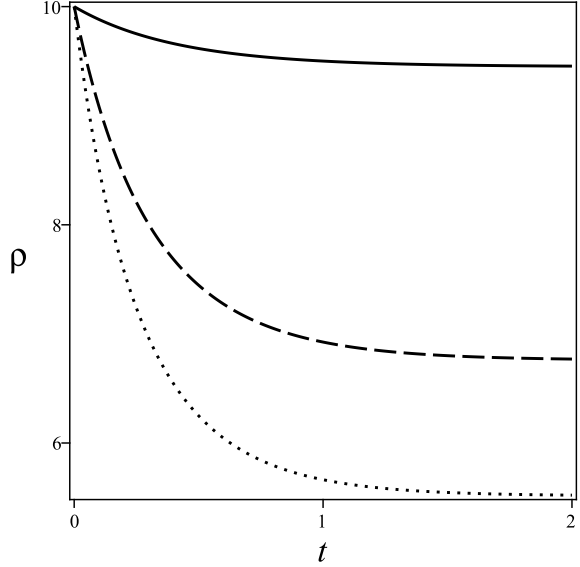}
\caption{Time-dependent dark energy density with $\alpha=0.5$ for
$n=0$ (dotted line), $n=0.2$ (solid line), $n=0.1$ (dashed line).}
\end{center}
\end{figure}

Now, instead of equation of state (1) we used the following equation
of state includes viscous coefficient,
\begin{equation}\label{s12}
\bar{p}\rightarrow
p=-\frac{A}{\rho^{\alpha}}-\sqrt{3}\zeta_{0}\rho^{n+\frac{1}{2}}.
\end{equation}
If $\zeta_{0}=-\frac{\sqrt{3}}{3}\gamma$ and $n=0.5$ then,
\begin{equation}\label{s13}
p=\gamma\varrho-\frac{A}{\rho^{\alpha}},
\end{equation}
where $\gamma$ is a positive constant, which is equation of state of
modified Chaplygin gas. It means that viscous generalized Chaplygin
gas may serves as modified Chaplygin gas. Therefore we can write
scale factor dependent energy density as the following,
\begin{equation}\label{s14}
\rho=\left[\frac{A}{1-\sqrt{3}\zeta_{0}}+\frac{C}{a^{3(1+\alpha)(1-\sqrt{3}\zeta_{0})}}\right]^{\frac{1}{1+\alpha}},
\end{equation}
where $C$ is an integration constant and may be choose as the
following,
\begin{equation}\label{s15}
C=-\frac{A}{1-\sqrt{3}\zeta_{0}}\frac{a(0)^{3(1+\alpha)(1-\sqrt{3}\zeta_{0})}}{\sqrt{3}\zeta_{0}}.
\end{equation}
It is clear that large values of the scale factor ($a\gg a(0)$) give
the following expression,
\begin{equation}\label{s16}
\rho\approx\left[\frac{A}{1-\sqrt{3}\zeta_{0}}\right]^{\frac{1}{1+\alpha}},
\end{equation}
which is corresponding to an empty universe.
\section{Conclusion}
In this paper we considered generalized Chaplygin gas which has
viscosity. Indeed we considered non-constant viscosity proportional
to $\rho^{n}$. Special case of $n=0$ and $\alpha=0.5$ yields to the
results of the Ref. [43] for $k=0$ (flat space). Two possibility of
positive and negative $n$ investigated by using plots of Fig. 1 and
Fig. 2. We found special case with $n=0.5$ and
$\zeta_{0}=-\frac{\sqrt{3}}{3}\gamma$, changed equation of state of
viscous generalized Chaplygin gas to modified Chaplygin gas.
Therefore, analogous to modified Chaplygin gas we obtained energy
density in terms of scale factor and saw that $\zeta_{0}=0$ limit of
the equation (14) reduced to the equation (8). Hence we expect that
this model will be stable. However obtaining solution for arbitrary
$n$ left for future study. Also It is interesting to consider
another dependents of viscosity such as $\zeta(t)$.

\end{document}